# The shape of the brain's connections is predictive of cognitive performance: an explainable machine learning study

Yui Lo[1,2,4], Yuqian Chen[1,2,4], Dongnan Liu[4], Wan Liu[5], Leo Zekelman[2,8], Jarrett Rushmore[3,6], Fan Zhang[7], Yogesh Rathi[1,2], Nikos Makris[1,3], Alexandra J. Golby[1,2], Weidong Cai[4], and Lauren J. O'Donnell[1,2,9]

[1] Harvard Medical School, Boston, USA
[2] Brigham and Women's Hospital, Boston, USA
[3] Massachusetts General Hospital, Boston, USA
[4] The University of Sydney, Sydney, Australia
[5] Beijing Institute of Technology, Beijing, China
[6] Boston University, Boston, USA
[7] University of Electronic Science and Technology of China, Chengdu, China
[8] Harvard University, Boston, USA
[9] Harvard-MIT Health Sciences and Technology, Cambridge, USA

Corresponding authors: odonnell@bwh.harvard.edu , zhangfanmark@gmail.com and tom.cai@sydney.edu.au

## Acknowledgments

We gratefully acknowledge funding provided by the following grants: National Institutes of Health (NIH) grants R01MH132610, R01MH125860, R01MH119222, R01NS125307, R01NS125781, and R21NS136960. FZ is in part supported by the National Key R&D Program of China (No. 2023YFE0118600), and the National Natural Science Foundation of China (No. 62371107). This work is also supported by The University of Sydney International Scholarship and Postgraduate Research Support Scheme.

## Conflicts of Interest

The authors declare no conflict of interest.

## Data Availability

The Human Connectome Project minimally preprocessed young adult dataset imaging data and associated NIH Toolbox measures are publicly available at https://db.humanconnectome.org/. The ORG tractography atlas is publicly available at http://dmri.slicer.org/atlases/, and code to apply the atlas is publicly available at https://github.com/SlicerDMRI/whitematteranalysis. All code developed for our experiments will be publicly available at https://github.com/SlicerDMRI/TractShapeCognition.

## IRB Statement

The creation of the WU-Minn HCP dataset was approved by the institutional review board of Washington University in St. Louis (IRB #201204036)

# Abstract

**Introduction**: The shape of the brain's white matter connections is relatively unexplored in diffusion MRI (dMRI) tractography analysis. While it is known that tract shape varies in populations and across the human lifespan, it is unknown if the variability in dMRI tractography-derived shape may relate to the brain's functional variability across individuals.

**Methods**: This work explores the potential of leveraging tractography fiber cluster shape measures to predict subject-specific cognitive performance. We implement two machine learning models (1D-CNN and LASSO) to predict individual cognitive performance scores. We study a large-scale database from the Human Connectome Project Young Adult study (n=1065). We apply an atlas-based fiber cluster parcellation (953 fiber clusters) to the dMRI tractography of each individual. We compute 15 shape, microstructure, and connectivity features for each fiber cluster. Using these features as input, we train a total of 210 models (using five-fold cross-validation) to predict 7 different NIH Toolbox cognitive performance assessments. We apply an explainable AI technique, SHAP (SHapley Additive exPlanations), to assess the importance of each fiber cluster for prediction.

**Results**: Our results demonstrate that fiber cluster shape measures are predictive of individual cognitive performance. The studied shape measures, such as irregularity, diameter, total surface area, volume, and branch volume, are generally as effective for prediction as traditional microstructure and connectivity measures. The 1D-CNN model generally outperforms the LASSO method for prediction. Further interpretation and analysis using SHAP values from the 1D-CNN suggest that fiber clusters with features highly predictive of cognitive ability are widespread throughout the brain, including fiber clusters from the superficial association, deep association, cerebellar, striatal, and projection pathways.

**Conclusion**: This study demonstrates the strong potential of shape descriptors to enhance the study of the brain's white matter and its relationship to cognitive function.

**Key Points**:

1. We investigated if white matter shape could predict cognitive performance
2. Using explainable machine learning, we trained 210 models to predict cognitive performance
3. Most shape measures are as predictive as microstructure or connectivity information
4. Results suggest the shape of the brain's fiber tracts is important for future study

# 1. Introduction

Diffusion magnetic resonance imaging (dMRI) tractography is an in-vivo imaging method to map white matter connections based on the water diffusion in brain tissue [Basser et al., 2000]. dMRI tractography produces sequences of points called streamlines that estimate the course of white matter connections. Streamlines can be organized into anatomical fiber tracts or finely parcellated into fiber clusters, enabling quantitative measures of white matter microstructure, connectivity, and shape [Zhang et al., 2022a]. These measures allow the study of the brain's white matter in health and disease and its relationship to non-imaging phenotypes, such as individual cognitive performance [Forkel et al., 2022]. Recent studies have demonstrated the prediction of non-imaging phenotypes from microstructure and connectivity measures at the fiber tract and fiber cluster levels [Chamberland et al., 2021; Chen et al., 2024a; Xiao et al., 2021; Zekelman et al., 2022]. These studies have employed regression or deep learning models to output a subject-specific non-imaging phenotype score. However, these studies mainly investigated traditional dMRI tractography measures, such as the number of streamlines (NoS), fractional anisotropy (FA), or mean diffusivity (MD). The NoS relates to the geometry of the connection but is usually thought of as a proxy for its connectivity or connection "strength" [Zhang et al., 2022a], while the FA and MD microstructure measures quantify the anisotropy and magnitude of water diffusion and are sensitive to a variety of tissue properties [Beaulieu, 2009; Jones et al., 2013].

While these microstructure and connectivity features are highly informative, they ignore many potentially important morphometric or shape characteristics of the brain's connections. Recently, shape measures, including length, area, volume, and other metrics, were applied to study the morphometry of the brain's association fiber tracts [Yeh, 2020]. Other research has shown that shape measures vary across the lifespan [Lebel et al., 2012; Schilling et al., 2023a; Schilling et al., 2023b]. While these studies have demonstrated that measures of fiber tract shape have both population and lifespan variability, it is unknown to what extent the dMRI tractography shape variability may relate to the brain's functional variability across individuals. Our recent preliminary work has shown that fiber cluster shape features have the potential to predict individual language functional performance [Lo et al., 2024a]. However, it is not yet known if dMRI tractography shape features are predictive of different cognitive abilities, which shape features may be the most predictive, and how their predictive performance may compare to more traditional features such as NoS and FA. The answers to these questions may help guide further research into the shape of the brain's connections.

To begin to address these questions, the main contribution of this work is to explore the potential of a comprehensive set of fiber cluster shape measures for the prediction of multiple domains of individual cognitive performance. We propose a computationally intensive data-driven strategy of training many (210) machine learning models to predict individual cognitive performance using multiple input microstructure, connectivity, and shape measures. By comparing the models' performance, we aim to achieve a broad, data-driven assessment of the potential of shape for studying the brain's white matter connections and their relationship to human brain functional performance.

# 2. Materials and Methods

## 2.1 Overview

Our overall strategy is as follows (Figure 1). We first perform a fine parcellation of the white matter into fiber clusters (Section 2.2.1), which have been shown to have improved power to predict human traits in comparison with traditional connectome matrices [Liu et al., 2023a]. We then calculate shape measures (12 features described below in Section 2.3.2), microstructure features (FA and MD), and NoS features for each fiber cluster for all subjects. Next, we use these features as inputs to train machine learning models (CNN and LASSO) that are known to successfully predict non-imaging phenotypes from tractography data [Chen et al., 2022; Cui and Gong, 2018; Feng et al., 2022; He et al., 2022; Liu et al., 2023b; Lo et al., 2024a; Wen et al., 2019]. We train the models (Section 2.4) to predict individual cognitive performance, where we use a testbed of 7 NIH Toolbox assessments (described below in Section 2.2.2). Finally, the predictive performance of the models is compared (Section 2.5) to assess the utility of the input measures for predicting cognitive performance. We implement an explainable module (Section 2.6) to describe the contribution of each fiber cluster measure to the prediction, and we provide example visualizations of the predictive white matter anatomy.

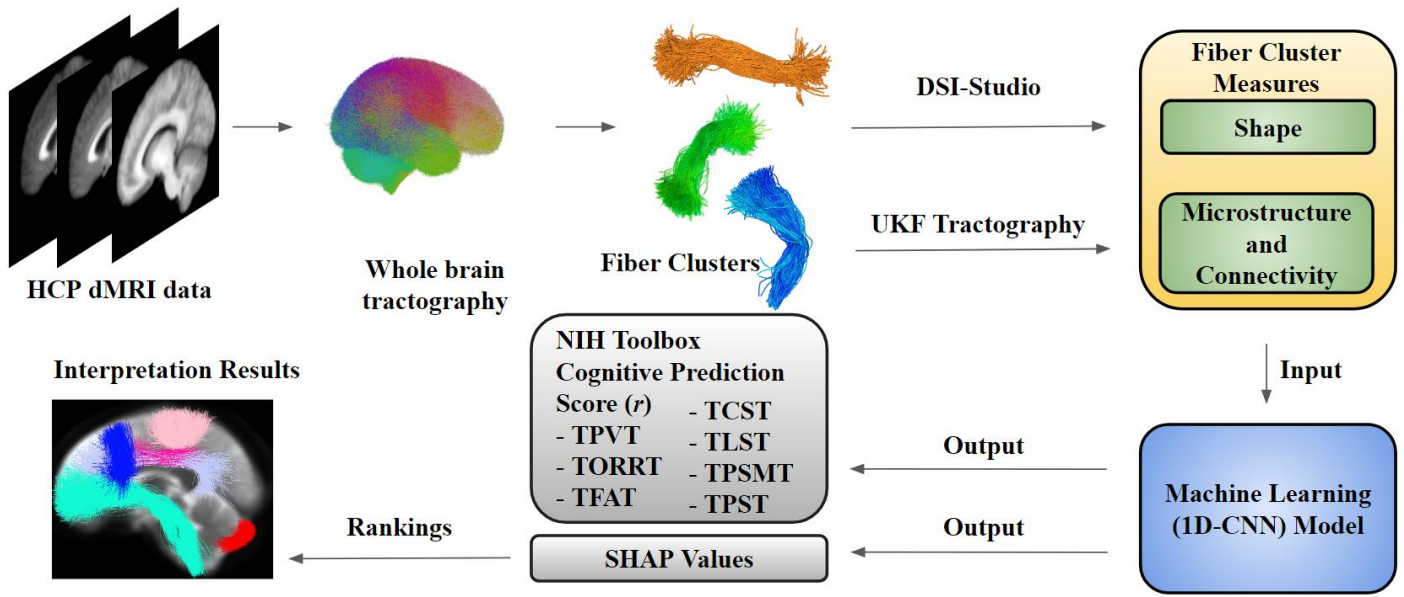

**Figure 1**. The overall data-driven pipeline of our work.

## 2.2 Study Material

### 2.2.1 Dataset, Tractography, and Fiber Clustering Parcellation

In this work, we study the brain connections of 1065 healthy young adults (575 females and 490 males, mean age 28.7 years) from the Human Connectome Project minimally preprocessed young adult dataset (HCP-YA) [Van Essen et al., 2012; Van Essen et al., 2013]. We used pre-computed HCP-YA tractography as employed in recent work [Chen et al., 2023; Zekelman et al., 2022]. Whole brain tractography was performed with a multi-tensor unscented Kalman filter (UKF) tractography method [Reddy and Rathi, 2016][1] that is highly consistent across the human lifespan, across test-retest scans, across disease states, and across different acquisitions [Zhang et al., 2018c; Zhang et al., 2019]. UKF is effective for reconstructing anatomical somatotopy [He et al., 2023]. The UKF method estimates a tissue microstructure model during fiber tracking, leveraging prior information from the previous tracking step to increase model fitting stability. UKF enables the estimation of tract-specific microstructural measures using the first tensor, which models the traced tract. Each subject's whole brain tractography was automatically parcellated into fiber clusters using whitematteranalysis [Zhang et al., 2018c]. This atlas-based machine learning method is consistent across the human lifespan, across test-retest scans, across disease states including brain tumors, and across different acquisitions [Zhang et al., 2018c][2] with high test-retest reproducibility [Zhang et al., 2019]. Fiber clusters are known to provide a compact vector representation of the connectome with improved power to predict human traits [Liu et al., 2023a; Zhang et al., 2018a], enabling a variety of downstream analyses [Chen et al., 2023; Gabusi et al., 2024; Xue et al., 2024; Zhang et al., 2018b]. The whitematteranalysis method parcellates robustly by using a spectral embedding of streamlines, a machine-learning technique that takes into account the variability across subjects [O'Donnell and Westin, 2007]. Fiber clusters were defined in the whitematteranalysis package using the O'Donnell Research Group (ORG) fiber cluster atlas [Zhang et al., 2018c]. The ORG atlas was anatomically curated to organize fiber clusters into anatomically labeled fiber tracts and to enable the automatic removal of anatomically inaccurate (false positive) fiber clusters. For each subject in the current study, the whitematteranalysis processing provides 953 fiber clusters, anatomically categorized into 58 deep white matter anatomical tracts, including association,

---

cerebellar, commissural, and projection tracts, and 16 superficial tract categories based on location within different brain lobes.

## 2.2.2 NIH Toolbox Cognitive Performance Assessments

In this work, we investigate the potential of shape measures for predicting non-imaging phenotypes. As a testbed for prediction, we choose seven non-imaging cognitive phenotypes that span the domains of language, executive function and attention, working and episodic memory, and processing speed (Table 1). These phenotypes are from the NIH Toolbox, a standard battery for neurobehavioral measurement [Hodes et al., 2013]. Specifically, we use the NIH Toolbox age-adjusted cognitive performance assessments provided by HCP-YA [Weintraub et al., 2013]. These assessments (Table 1) are the NIH Toolbox Picture Vocabulary Test (TPVT), the NIH Toolbox Oral Reading Recognition Test (TORRT), the NIH Toolbox Flanker Inhibitory Control and Attention Test (TFAT), the NIH Toolbox Dimensional Change Card Sort Test (TCST), the NIH Toolbox List Sorting Working Memory Test (TLST), the NIH Toolbox Picture Sequence Memory Test (TPSMT), and the NIH Toolbox Pattern Comparison Processing Speed Test (TPST).

Table 1: Overview of NIH Toolbox cognitive assessments used in this study.

| Category | Assessment | Abbreviation | Description | Citation |
|---|---|---|---|---|
| Language | NIH Toolbox Picture Vocabulary Test | TPVT | tests the ability to select the picture that corresponds to a spoken word | [Gershon et al., 2014] |
| | NIH Toolbox Oral Reading Recognition Test | TORRT | assesses the ability to pronounce individual printed words and identify letters | |
| Executive function and attention | NIH Toolbox Flanker Inhibitory Control and Attention Test | TFAT | assesses the ability to focus on a particular stimulus while inhibiting attention to the stimuli flanking it | [Zelazo et al., 2013] |
| | NIH Toolbox Dimensional Change Card Sort Test | TCST | assesses the flexibility to match a series of picture pairs to a target picture | |
| Working and episodic memory | NIH Toolbox List Sorting Working Memory Test | TLST | assesses working memory to recall and sequence different stimuli that are presented visually and via audio | [Tulsky et al., 2014] |
| | the NIH Toolbox Picture Sequence Memory Test | TPSMT | assesses episodic memory to remember a sequence of pictures of objects and activities | [Bauer et al., 2013] |
| Processing speed | NIH Toolbox Pattern Comparison Processing Speed Test | TPST | measures the ability to quickly determine whether two stimuli are the same or different | [Carlozzi et al., 2015] |

## 2.3 Fiber Cluster Measures

### 2.3.1 Microstructure and Connectivity Measures

We study 3 widely used FA, MD, and NoS measures, which are calculated for each of the 953 fiber clusters from all subjects. These popular measures are informative in the study of the relationship between the brain's fiber tracts and cognition [Chen et al., 2024a; Forkel et al., 2022; Zekelman et al., 2022]. FA measures how much a tensor deviates from a sphere, quantifying the anisotropy of water molecule diffusion. MD reflects the total amount of water diffusion within the tissue, providing an average measure of the diffusion in all directions. NoS can be considered a geometry or shape measure but is popularly considered a "connectivity" measure and is indirectly related to structural connectivity "strength" [Qi et al., 2015; Zhang et al., 2022a].

### 2.3.2 Shape Measures

We study 12 fiber cluster shape measures that are considered to provide a detailed and comprehensive shape analysis of tractography [Yeh, 2020]. These measures include length, span, volume, diameter, curl, elongation, trunk volume, branch volume, total surface area, total area of end regions, total radius of end regions, and irregularity (Figure 2). These shape measures are computed for all 953 fiber clusters from all subjects by applying the software package DSIStudio [Yeh, 2020][3].

Full definitions of the shape measures can be seen in [Yeh, 2020], and we include a brief description of each measure here. Length is the average streamline length of the fiber cluster in mm. Span is the distance between the two ends of the fiber cluster in mm. Volume is defined as the volume of voxels occupied by the fiber cluster in mm$^3$, where DSIStudio creates a synthetic volume of isotropic voxels for use in computations. Diameter is estimated from the length and volume shape measures, under the assumption that the fiber cluster takes the form of a cylinder; the diameter is considered to represent the average fiber cluster diameter in mm. Curl is defined as the length divided by the span to approximate the overall curve (note this is also known as the u-ratio in other literature, e.g. [Nie et al., 2024]). Elongation is defined as the fiber cluster length divided by the diameter. The trunk volume is the volume of the streamlines connecting the fiber cluster's two largest end regions. The branch volume is the remaining volume (the difference between the total and trunk volume). The surface area is estimated by counting the outermost or surface voxels occupied by the fiber cluster, then multiplying by the area of one side of a voxel in mm$^2$. The total area of end regions is computed as above using the surface voxels of the cluster endpoints, which are also used in estimating the total radius of end regions using a circular model. Finally, irregularity is computed by dividing the fiber cluster surface area by the surface area of a cylinder with the same diameter and length; thus, irregularity quantifies how different the shape is from an idealized cylinder.

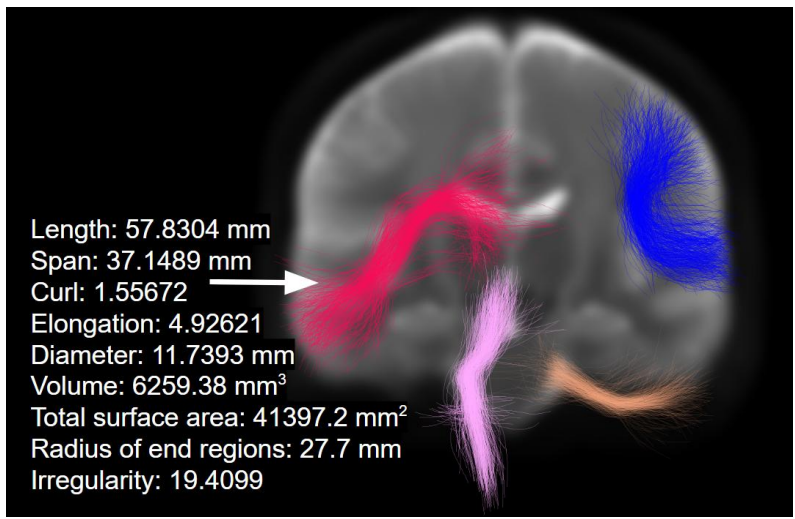

**Figure 2.** Four examples of individual white matter connections (fiber clusters) extracted from the entire white matter of the human brain using a fiber clustering approach. Example shape descriptors are shown for the red fiber cluster.

---

[3] https://dsi-studio.labsolver.org/

## 2.4 Machine Learning Prediction

To assess whether the fiber cluster shape measures help predict individual cognitive performance, we employ two machine learning prediction methods: one deep learning method and one traditional method.

First, we implement a deep learning method using a 1-D CNN model. This approach has previously successfully predicted non-imaging phenotypes from tractography data [He et al., 2022; Liu et al., 2023b; Lo et al., 2024a]. Specifically, the input of each model is one of the 15 shape, microstructure, or connectivity measures measured from all 953 fiber clusters. The model output is a predicted cognitive performance score (one of the seven described above). In technical details, the model's architecture is designed as follows. The model consists of a feature extractor and a classifier. The model is trained using mean squared error loss by comparing its output with the ground truth. The feature extractor contains three convolutional blocks, each comprising a 1D convolutional layer (kernel size=5, kernel number=64, stride=1), a batch normalization layer with a momentum of 0.5, a ReLU activation layer, and a dropout layer for regularization. The classifier comprises two fully connected layers; the first layer reduces the input dimension from 512 to 128, followed by batch normalization with a momentum of 0.5, a ReLU activation layer, and dropout. The final layer maps the output to the predicted score of the cognitive performance. This deep multi-layered architectural design aims to effectively learn from the input measure to improve the overall prediction of the cognitive performance score.

For comparison, we implement LASSO (Least Absolute Shrinkage and Selection Operator) regression [Feng et al., 2022; Tibshirani, 2018], a conventional machine learning method that leverages variable selection and regularization. LASSO regression has been used successfully to predict cognitive performance from brain connectome and fiber tract data [Chen et al., 2022; Cui and Gong, 2018; Feng et al., 2022; Wen et al., 2019].

Overall, in this project, a model is trained to predict each cognitive assessment from each input feature. This results in a total of 210 machine learning models were trained, including the 105 (15 input features x 7 cognitive assessments) 1D-CNN models and the 105 (15x7) LASSO models. This approach ensures that each model focuses on a specific input-prediction pair, reducing the complexity of training. This experimental design enables the comparison of prediction performance across all input features. 5-fold cross-validation was used for training each model (see Section 2.7 for more details).

## 2.5 Statistical Analysis

We conduct a statistical analysis of the prediction performance results. The correlation coefficient (Pearson's *r*) is not normally distributed. Therefore, we transform the results from all five folds to z-scores using the widely applied Fisher's r-to-z transformation [Chen et al., 2024a; Keller et al., 2011; Shen et al., 2015; Tobyne et al., 2018]. To compare the performance of different models, we employ a one-way repeated measures Analysis of Variance (ANOVA) of the z-transformed correlation coefficients, followed by post hoc pairwise comparisons using paired t-tests between each of the different models.

## 2.6 Interpretation with Explainable AI

To provide potential insight into how the shape, microstructure, and connectivity features inform the prediction of individual cognitive performance, we implement a widely used explainable AI approach, SHAP (SHapley Additive exPlanations). SHAP assigns each feature an importance value for a particular prediction [Lundberg and Lee, 2017]. The computation of SHAP values is based on a coalition game theory approach, providing a clear and interpretable explanation to improve the trustworthiness of the interpreted model. The SHAP approach has been widely applied for the interpretation of brain features that affect the prediction of individual intelligence, memory, language [Azevedo et al., 2019], and brain age [Ballester et al., 2023; Scheda and Diciotti, 2022; Sun et al., 2022]. In this study, the objective of leveraging SHAP interpretation is to understand the model's insights into the brain's white matter connections and how their microstructure, connectivity, and shape features may relate to cognitive performance.

To interpret the contribution of each of the 953 fiber clusters, we evaluate which fiber clusters most impact the prediction results. To compute this, for each of the 7 NIH toolbox assessment prediction models, we calculate each cluster's mean SHAP value across the 5 folds. Then, we rank the clusters for each model according to SHAP value. Finally, to assess the overall impact on prediction results across models, we average across the 7 different rankings to get a mean ranking for each cluster. We sort clusters according to mean rankings in ascending order to evaluate the most impactful fiber clusters for the overall prediction of cognitive assessment scores.

## 2.7 Implementation Details

We optimize our 1D-CNN model with the Stochastic Gradient Descent (SGD) algorithm [Zou and Hastie, 2005]. The SGD optimizer is instantiated with a learning rate of 0.01 and trained on 200 epochs with batch sizes of 8, and the number of workers is set to 10. All experiments in this work were performed on an NVIDIA RTX A5000 GPU using PyTorch 1.8.1 [Paszke et al., 2019] across a 5-fold cross-validation to ensure reproducibility and consistency in model training and evaluation. We use the sklearn python package [Pedregosa et al., 2011] for the LASSO regression model with alpha set to 1.

All 105 1D-CNN model training experiments were run with the same fine-tuned hyperparameters. All 105 1D-CNN and 105 LASSO model training experiments were run with the same data split of 80% training data and 20% testing data across the five folds.

# 3. Results

## 3.1 Evaluation Metric

We employ the Pearson correlation coefficient (Pearson's *r*) *[Sedgwick, 2012]* to evaluate model performance. This metric is widely applied in neurocognitive performance prediction [Chen et al., 2020; Feng et al., 2022; Gong et al., 2021; Jeong et al., 2021; Kim et al., 2021; Rasero et al., 2021; Tian and Zalesky, 2021; Wu et al., 2023; Xue et al., 2024]. Pearson's *r* measures the strength and direction (positive or negative) of the linear association between two variables.

## 3.2 Prediction Performance Results

Table 2 gives the model performance (*r*) for the 105 (15x7) 1D-CNN models trained using the 15 input features (microstructure, connectivity, and shape) to predict the 7 output NIH toolbox assessment scores. The top-performing model in each column is shown in bold and represents the highest *r* for predicting a particular NIH toolbox assessment score. Other models with statistically equivalent performance to the top-performing model are shown in italics.

Overall, the results shown in Table 2 suggest that many shape features are as predictive as the widely used microstructure and connectivity features. Many features exhibit statistically equivalently high performance (in bold and italics), as shown in each column of Table 2. On average (rightmost column of Table 2), the highest-performing features include microstructure (FA and MD), connectivity (NoS), and the top-performing shape measures of Irregularity, Diameter, Total Surface Area, Branch Volume, Volume, Surface Area of End Regions, and Elongation. These features are similarly predictive overall on average across the seven prediction tasks. In fact, the overall highest-performing measure is a shape measure, Irregularity (though its performance is not significantly different from the other highest-performing measures). While the highest-performing shape measures have similar performance to the microstructure and connectivity features, it can be noted that several shape measures do have lower performance on average in this experiment (Radius of End Regions, Curl, Span, Length, and Trunk Volume). In this experiment, the least informative feature is Trunk Volume.

Table 2: Pearson correlation coefficients ($r$) of the 1DCNN for predicting seven NIH Toolbox cognitive assessments. Values are the mean and standard deviation across 5-fold cross-validation. The best result for each column is bolded. Results that are not significantly different from the best result in each column are in italics.

| | Language | | Executive function | | Memory | | Processing speed | |
|---|---|---|---|---|---|---|---|---|
| | TPVT | TORRT | TFAT | TCST | TLST | TPSMT | TPST | Average $r$ |
| **Microstructure** | | | | | | | | |
| FA | *0.314 ± 0.08* | *0.281 ± 0.06* | *0.132 ± 0.03* | 0.053 ± 0.03 | **0.239 ± 0.03** | *0.128 ± 0.03* | *0.135 ± 0.03* | *0.183 ± 0.1* |
| MD | *0.263 ± 0.02* | 0.207 ± 0.04 | *0.181 ± 0.04* | 0.076 ± 0.05 | *0.209 ± 0.06* | **0.147 ± 0.04** | *0.134 ± 0.08* | *0.174 ± 0.07* |
| **Connectivity** | | | | | | | | |
| NoS | *0.328 ± 0.07* | **0.294 ± 0.05** | *0.179 ± 0.05* | **0.148 ± 0.07** | *0.229 ± 0.07* | *0.096 ± 0.04* | 0.116 ± 0.04 | *0.199 ± 0.09* |
| **Shape** | | | | | | | | |
| Irregularity | *0.312 ± 0.06* | *0.281 ± 0.07* | **0.183 ± 0.06** | *0.103 ± 0.04* | *0.224 ± 0.04* | *0.115 ± 0.03* | *0.177 ± 0.05* | **0.2 ± 0.08** |
| Diameter | **0.335 ± 0.06** | *0.289 ± 0.06* | *0.166 ± 0.06* | *0.119 ± 0.06* | *0.199 ± 0.02* | 0.09 ± 0.03 | *0.173 ± 0.03* | *0.196 ± 0.09* |
| Total Surface Area | *0.318 ± 0.05* | *0.277 ± 0.07* | *0.15 ± 0.05* | *0.109 ± 0.02* | *0.192 ± 0.03* | *0.106 ± 0.04* | **0.188 ± 0.05** | *0.191 ± 0.09* |
| Branch Volume | *0.316 ± 0.09* | *0.283 ± 0.07* | *0.134 ± 0.05* | *0.126 ± 0.05* | *0.209 ± 0.03* | 0.1 ± 0.01 | *0.155 ± 0.06* | *0.189 ± 0.09* |
| Volume | *0.325 ± 0.07* | *0.253 ± 0.09* | 0.122 ± 0.05 | *0.126 ± 0.04* | *0.215 ± 0.03* | *0.102 ± 0.03* | *0.175 ± 0.06* | *0.188 ± 0.09* |
| Surface Area of End Regions | *0.34 ± 0.06* | *0.276 ± 0.05* | *0.13 ± 0.03* | 0.082 ± 0.06 | *0.227 ± 0.01* | *0.072 ± 0.02* | *0.139 ± 0.02* | *0.181 ± 0.11* |
| Elongation | 0.245 ± 0.05 | 0.212 ± 0.03 | *0.157 ± 0.06* | *0.143 ± 0.04* | *0.192 ± 0.06* | 0.054 ± 0.04 | *0.159 ± 0.03* | *0.166 ± 0.07* |
| Radius of End Regions | 0.229 ± 0.05 | 0.24 ± 0.04 | *0.139 ± 0.03* | *0.11 ± 0.04* | 0.155 ± 0.04 | *0.109 ± 0.05* | *0.162 ± 0.05* | 0.164 ± 0.06 |
| Curl | 0.189 ± 0.08 | *0.208 ± 0.10* | *0.139 ± 0.06* | *0.067 ± 0.03* | 0.153 ± 0.06 | *0.115 ± 0.05* | *0.146 ± 0.10* | 0.145 ± 0.05 |
| Span | 0.253 ± 0.04 | 0.204 ± 0.07 | 0.124 ± 0.04 | *0.084 ± 0.03* | *0.176 ± 0.06* | 0.057 ± 0.04 | 0.091 ± 0.05 | 0.141 ± 0.08 |
| Length | 0.214 ± 0.07 | 0.185 ± 0.07 | 0.087 ± 0.05 | *0.093 ± 0.04* | 0.131 ± 0.08 | *0.071 ± 0.04* | 0.092 ± 0.04 | 0.125 ± 0.06 |
| Trunk Volume | 0.126 ± 0.072 | 0.104 ± 0.08 | 0.076 ± 0.05 | *0.074 ± 0.07* | 0.083 ± 0.05 | *0.095 ± 0.04* | 0.108 ± 0.05 | 0.095 ± 0.02 |

For a complementary assessment of the utility of each feature to predict subject-specific cognitive performance, Table 3 summarizes the prediction performance ($r$) across the LASSO and CNN models. (Note that for the 1D-CNN model, these values are the same as the values in the rightmost column of Table 2. Also, more details on the LASSO results are provided in Supplementary Table S1. Overall, the 1D-CNN method generally outperforms the LASSO method. However, the LASSO method has better worst-case performance and provides improved prediction performance for the lower-performing shape measures (Curl, Span, Length, Trunk Volume), suggesting their potential utility. The statistical analysis (repeated measures ANOVA) of the LASSO results finds no significant performance difference across the models trained with the 15 features (p=0.37), suggesting that all investigated microstructure, connectivity, and shape features have utility for the study of the brain.

Table 3: Pearson correlation coefficients ($r$) of the LASSO and 1D-CNN models. Values are the mean and standard deviation across the 7 models trained to predict the 7 NIH Toolbox cognitive assessments. The best result for each column is bolded. In each column, results that are not significantly different from the best result are in italics.

| | LASSO Regression AVG | CNN AVG |
|---|---|---|
| **Microstructure** | | |
| FA | *0.157 ± 0.05* | *0.183 ± 0.1* |
| MD | *0.159 ± 0.04* | *0.174 ± 0.07* |
| **Connectivity** | | |
| NoS | **0.18 ± 0.03** | *0.199 ± 0.09* |
| **Shape** | | |
| Irregularity | *0.174 ± 0.03* | **0.2 ± 0.08** |
| Diameter | *0.148 ± 0.07* | *0.196 ± 0.09* |
| Total surface area | *0.171 ± 0.03* | *0.191 ± 0.09* |
| Branch Volume | *0.152 ± 0.07* | *0.189 ± 0.09* |
| Volume | *0.174 ± 0.03* | *0.188 ± 0.09* |
| Surface area of end regions | *0.155 ± 0.05* | *0.181 ± 0.11* |
| Elongation | *0.175 ± 0.03* | *0.166 ± 0.07* |
| Radius of end regions | *0.149 ± 0.03* | 0.164 ± 0.06 |
| Curl | *0.153 ± 0.04* | 0.145 ± 0.05 |
| Span | *0.148 ± 0.04* | 0.141 ± 0.08 |
| Length | *0.145 ± 0.04* | 0.125 ± 0.06 |
| Trunk Volume | *0.159 ± 0.03* | 0.095 ± 0.02 |
| **Average** | 0.163 ± 0.01 | 0.169 ± 0.03 |

## 3.3 Explainable Interpretation

In this section, we provide potentially informative visualizations of the brain fiber clusters whose input features are most predictive of individual cognition. We focus on the interpretation of the best-performing models ($r > 0.18$ on average across all NIH Toolbox assessments). This includes the highest-performing 1D-CNN models (Table 3). We provide visualizations of the top 10 clusters for predicting individual cognitive performance using each input feature, along with the indices and anatomical labels of the clusters (Figure 3).

The top 10 clusters for different input features can be observed to belong to multiple different anatomical brain fiber tracts, including different types of tracts (superficial, association, cerebellar, striatal, and projection). However, commissural clusters (e.g., corpus callosum) are not identified in the top 10 most predictive clusters for any input feature in this experiment. In Figure 3 it can be observed that, for example, the top 10 most predictive fiber clusters for FA are all association fiber clusters, including 9 superficial white matter fiber clusters (one is bilateral, superficial frontal cluster 604) and one left uncinate fiber cluster. As another example, the top 10 predictive fiber clusters for Total Surface Area include 4 superficial white matter fiber clusters, 3 association clusters in the middle longitudinal fasciculus (bilateral cluster 691) and left cingulum, 2 intracerebellar parallel tract clusters (bilateral cluster 513), and a left striato-frontal projection cluster. The predictive clusters for the top-performing input feature, Irregularity, span multiple fiber tract categories including 2 deep association fiber clusters (left cingulum), 4 superficial association fiber clusters, 2 cerebellar clusters (bilateral parallel tract cluster 516), 1 striatal cluster (left external capsule), and 1 projection cluster (left corona radiata frontal). Overall, this result suggests that fiber clusters with features predictive of cognitive ability are widespread throughout the brain and that different features capture different informative aspects of the white matter anatomy.

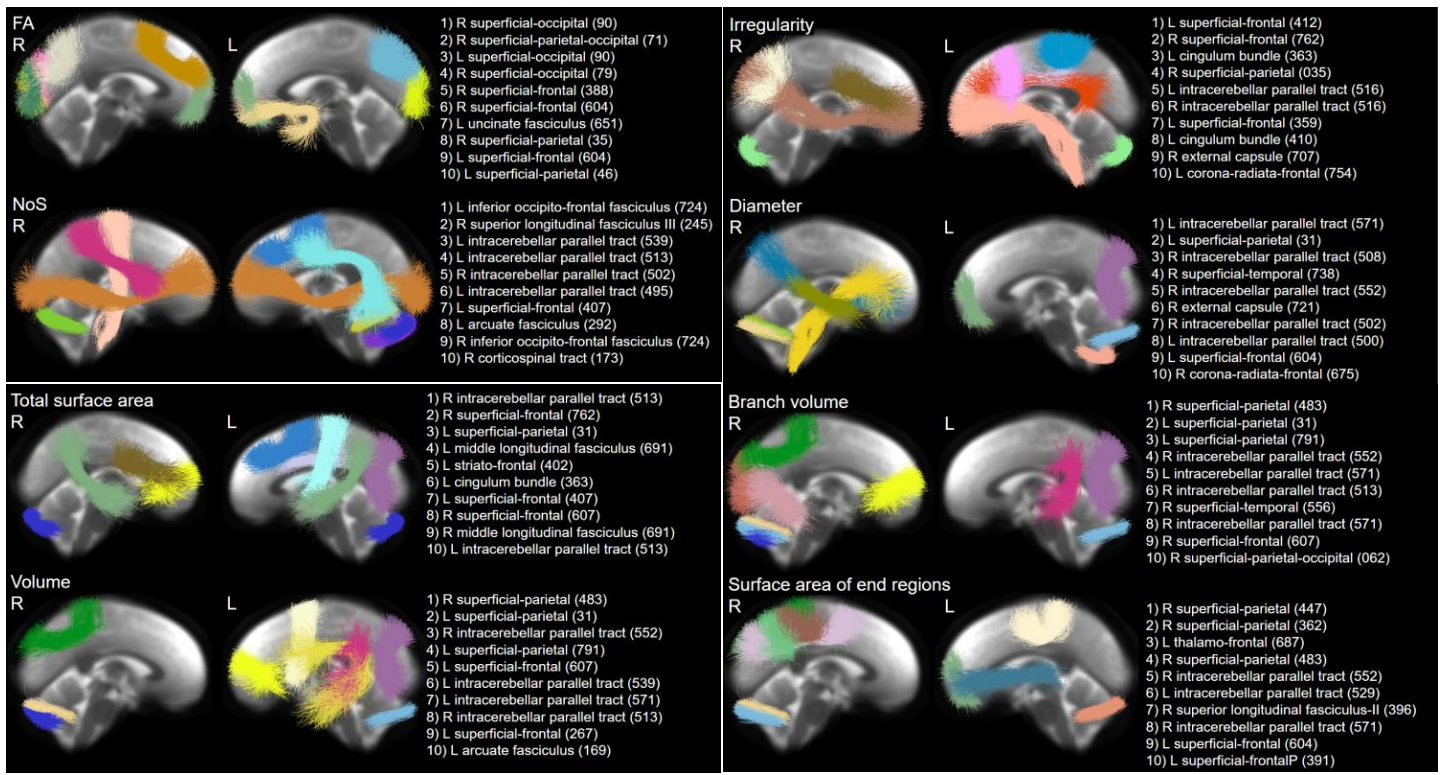

**Figure 3**. Visualization of the top ten most predictive fiber clusters for predicting individual cognitive performance using different input features. Fiber clusters are located within white matter fiber tracts, as listed to the right of the images (with fiber cluster atlas ID numbers provided in parentheses). Visualizations are performed using the fiber cluster atlas [Zhang et al., 2018c].

# 4. Discussion

This work has explored the potential of leveraging fiber cluster shape measures to predict subject-specific cognitive performance. We employed a data-driven strategy leveraging a large set of shape characteristics and assessments of individual cognitive performance. We applied two machine learning methods, 1D-CNN and LASSO regression. While both machine learning methods demonstrated that the proposed shape measures performed well in predicting individual cognitive performance, the 1D-CNN was more predictive on average. We introduced the widely used explainable approach, SHAP, to identify highly predictive fiber clusters. Our experimental results on a large-scale dataset of 1065 subjects demonstrated the effectiveness of shape measurements for individual cognitive performance prediction and identified white matter fiber clusters that highly contributed to the overall performance. Below, we discuss some detailed observations about our results in comparison with related studies in the literature.

Other groups have studied tractography shape measures over the past twenty years. Local shape measures, such as curvature and torsion, were proposed early in the field of tractography to describe local properties at points along individual streamlines [Batchelor et al., 2006; Corouge et al., 2004]. More recently, shape measures describing global properties of entire fiber tracts, such as the shape measures studied in this paper, have received increasing interest. Multiple research groups have applied the DSIStudio shape computation software [Yeh, 2020] to demonstrate that the shape of white matter fiber tracts varies in health and disease and across the lifespan [Linn et al., 2024; Schilling et al., 2022; Schilling et al., 2023a; Schilling et al., 2023b; Yang et al., 2024; Yin et al., 2023].

Several recent studies have assessed potential relationships between white matter fiber tract shape and various measures of brain function, with a mix of positive and negative findings. These studies have generally focused on particular anatomical fiber tracts and their relationship with the brain functions the tracts are expected to subserve (for a review of fiber tracts and functions, see [Forkel et al., 2022]). For example, the shape (radius and irregularity of end regions) of the corticospinal tract was significantly associated with the risk of postoperative motor complications in low-grade glioma patients [Yang et al., 2024]. In another example, the shape (volume and diameter) of the forceps minor of the corpus callosum was not associated with cognitive reappraisal, a mechanism for emotion regulation, in a healthy subjects dataset [Porcu et al., 2024]. In a recent preprint, a study of the HCP-YA found that functional language lateralization was not significantly associated with shape metrics (including length, span, curl, elongation, diameter, volume, and surface area) of eleven language-related white matter tracts [Andrulyte et al., 2024]. However, this study did not investigate assessments of language performance. Another recent study of the HCP-YA found a weak correlation of language performance assessments with a shape measure, the inferior end region coverage area of the frontal aslant tract [Linn et al., 2024]. Overall, it can be noted that the above-related work provides only weak evidence that fiber tract shape features may relate to cognitive performance in healthy individuals. However, an initial conference publication by our group demonstrated the strong potential of shape features for machine-learning-based prediction of language performance [Lo et al., 2024a]. In contrast to these related investigations, which studied a limited number of fiber tracts and cognitive assessments, the current study aimed to comprehensively evaluate multiple shape features, cognitive measures, and brain connections in the form of fiber clusters within fiber tracts. We sought to assess whether shape features predict cognitive abilities, which features are most predictive, and how their performance compares to traditional metrics like NoS and FA.

The results of the current study, which demonstrate that the investigated shape measures are predictive of cognitive performance, suggest that these shape measures are informative for studying the brain's white matter connections and their relationships to non-imaging phenotypes such as cognition. Furthermore, the results suggest that the study of shape features may be equally as informative as the study of traditional microstructure or connectivity features. Tables 1 and 2 provide clear evidence that the shape measures generally have similar predictive performance to traditional measures. While almost all shape measures perform reasonably well on average ($r > 0.1$), there is one clear exception where the Trunk Volume measure does not provide good performance when input to the 1D-CNN. The Trunk Volume was previously described as having low reliability in the test-retest sense [Yeh, 2020]. However, we note that the Trunk Volume measure performed well when using the LASSO models. Thus, our results suggest that all shape measures are informative not only about brain structure but also about the relationship between brain structure and brain function.

The application of explainable AI suggests that fiber clusters with features highly predictive of cognitive ability are widespread throughout the brain, including superficial association, deep association, cerebellar, striatal, and projection fiber clusters. The results further suggest that different features are able to capture different informative aspects of the white matter fiber cluster anatomy. As expected, some highly predictive fiber clusters (Figure 3) can be observed within fiber tracts expected to relate to aspects of cognition. For example, NoS and volume features of fiber clusters within the left arcuate fasciculus, classically critical for language function, are highly predictive of cognition. Similarly predictive are the total surface area and irregularity features of fiber clusters within the left cingulum bundle, a tract believed to be involved in multiple aspects of cognition [Forkel et al., 2022]. Many features extracted from the fiber clusters within the superficial white matter, which provides cognitively critical cortico-cortical communication, are predictive of cognition [Wang et al., 2022]. In Figure 3, it can also be observed that some clusters (e.g. superficial frontal cluster 607 and intracerebellar parallel tract cluster 571) are found to be highly predictive across several different features. This observation suggests that (as we already know) the shape features are not mathematically independent of each other. It also suggests that such consistently informative clusters may have the potential for future investigation in a more hypothesis-driven fashion.

The widespread nature of the predictive fiber clusters is in line with the understanding of cognition as arising from the interaction of multiple interconnected brain regions [Bressler and Menon, 2010; Mesulam, 1990; Uddin et al., 2019] and with a recent review demonstrating that fiber tracts are generally correlated with multiple cognitive domains [Forkel et al., 2022]. While the biological mechanisms linking white matter shape and cognitive performance are not fully understood, genetic and environmental factors influencing brain development may play a role [Luo et al., 2022]. Differences in white matter shape could also be linked to structural changes in adjacent gray matter regions, including increased gyrification of neocortical regions, a factor associated with cognitive ability [Gregory et al., 2016; Lamballais et al., 2020] but see [Mathias et al., 2020]. Overall, variability in shape features is likely to reflect numerous biological, environmental, genetic, and developmental factors that work in conjunction to influence the shape of an individual's brain and cognitive function.

We implemented the machine learning used in this paper using two popular methods previously shown to be successful for the prediction of non-imaging phenotypes. Our methodological choices enabled us to perform the current computationally intensive, data-driven study. However, there are several limitations. The 1D-CNN may overlook potential synergies between shape, microstructure, and connectivity features. In future work, algorithms such as multi-view learning [Sun, 2013; Wei et al., 2023; Xu et al., 2013] or ensembling learning [Gupta et al., 2017; Polikar, 2012; Sagi and Rokach, 2018; Zhang et al., 2022b] could extract information from hidden correlations between features that remain unexplored when trained independently. Additionally, 1D-CNNs have proven effective in capturing local patterns in data but are inherently limited when modeling global features, especially in the context of high-dimensional data. Future work could benefit from exploring more sophisticated deep models, such as conditional diffusion-based models [Lo et al., 2024b; Yao et al., 2023] or transformer-based models [Chen et al., 2024b; Lo et al., 2024a; Zhang et al., 2022b]. We implemented the explainable AI used in this paper using a popular method, SHAP, that is widely applied across different machine learning approaches. In our study, we did not rely solely on the empirical SHAP values, as the values are not always informative in isolation. Instead, we ranked the SHAP values and averaged them to focus on predictive performance for overall cognition. Other approaches for organizing SHAP values, such as clustering [Clement et al., 2024; Durvasula et al., 2022], may provide additional insight. Future work could also investigate other explainable AI techniques, such as activation-based methods [Zhou et al., 2016], attention mechanisms [Devlin et al., 2018; Dong et al., 2021], gradient-based approaches [Selvaraju et al., 2017], or alternative perturbation methods [Ribeiro et al., 2016b].

As one of the first studies to explore shape as a predictor of individual cognitive performance, this work has several key limitations and potential areas for future improvement. We investigated a large healthy young adult dataset. Future work may investigate the effectiveness of shape measures in other datasets, such as those from different acquisitions or across the lifespan in health and disease. Future validation in such independent datasets [Casey et al., 2018; Cetin-Karayumak et al., 2024; Ge et al., 2023; Marek et al., 2011] can enable the assessment of the robustness and generalizability of the predictive models. Such large datasets can also enable investigation of the impact of potentially confounding factors such as education and socioeconomic status, which may influence performance on cognitive assessments and measures of the white matter. Furthermore, we relied on a single tractography method and a single atlas-based definition of major fiber tracts [Zhang et al., 2018]. An in-depth comparison of tractography methods and white matter tract definitions [He et al., 2023; Schilling et al., 2021] remains an interesting future work to obtain a more robust understanding of the effectiveness of shape measures in different methodological scenarios. We also relied on a single method for interpretation, SHAP. In future work, combining or comparing additional model-agnostic interpretation approaches may offer alternative insights [Apley and Zhu, 2020; Ribeiro et al., 2016; Ribeiro et al., 2018]. In addition, it remains to be investigated if shape may predict other non-imaging phenotypes beyond NIH Toolbox assessments of cognition. Future work could explore other structure-function relationships, using additional non-imaging phenotypes and potentially more sophisticated machine learning and explainable AI approaches.

# 5. Conclusion

In this work, we investigated fiber cluster shape features, a novel approach for cognitive performance prediction. Using explainable deep learning, we demonstrated that the shape of the brain's connections is predictive of individual cognitive performance. We showed that shape can achieve better or equivalent prediction performance on cognitive prediction in comparison to conventional measurements of microstructure and connectivity. Our results suggest that all shape measures under study are relevant for the study of the brain's white matter connections and their relationships to non-imaging phenotypes such as cognition. Overall, this study demonstrates the potential of utilizing geometric shape descriptors of tractography fiber clusters to enhance the study of the brain's white matter and its relationship to various cognitive functions.